\documentclass[fleqn,usenatbib]{mnras}
\usepackage{newtxtext,newtxmath}

\usepackage[T1]{fontenc}

\DeclareRobustCommand{\VAN}[3]{#2}
\let\VANthebibliography\thebibliography
\def\thebibliography{\DeclareRobustCommand{\VAN}[3]{##3}\VANthebibliography}

\usepackage{graphicx}	% Including figure files
\usepackage{amsmath}	% Advanced maths commands

\title[Reverse shock emission from an off-axis top-hat jet]{Reverse Shock Emission in an Off-axis Top-hat Jet Model for Gamma-Ray Bursts}

\author[Senlin Pang et al.]{
Sen-Lin Pang$^{1}$ and Zi-Gao Dai$^{1,2}$\thanks{E-mail: daizg@ustc.edu.cn}
\\
$^{1}$Department of Astronomy, University of Science and Technology of China , Hefei 230026, China \\
$^{2}$School of Astronomy and Space Science, University of Science and Technology of China , Hefei 230026, China\\
}

\date{Accepted XXX. Received YYY; in original form ZZZ}

\pubyear{2015}

\begin{document}
\label{firstpage}
\pagerange{\pageref{firstpage}--\pageref{lastpage}}
\maketitle

\begin{abstract}
The afterglow of a gamma-ray burst (GRB) has been widely argued to arise from the interaction of  a relativistic outflow with its ambient medium. During such an interaction, a pair of shocks are generated: a forward shock that  propagates into the medium, and a reverse shock that propagates into the outflow. Extensive studies have been conducted on the emission from the forward shock viewed off-axis.
Furthermore, the observation of a reverse shock in an on-axis short GRB suggests that the reverse shock can produce an electromagnetic counterpart to a gravitational wave-detected merger. In this paper, we investigate the contribution of the reverse shock to the afterglow from a top-hat jet viewed off-axis, and apply our model to some short GRBs previously modeled by an off-axis emission. 
We employ the Markov Chain Monte Carlo (MCMC) method to get the model parameters (i.e.,  the jet's half-opeaning angle $\theta_j$, the viewing angle $\theta_\text{obs}$, the initial Lorentz factor $\Gamma_0$, and the isotropic energy $E_\mathrm{iso}$). 
Our model successfully reproduces off-axis afterglow emission without a structured jet. 
In addition, our calculations suggest that the reverse shock may produce a prominent feature in an early afterglow, which can be potentially observed in an orphan optical afterglow.
\end{abstract}


%%%%%%%%%%%%%%%%%%%%%%%%%%%%%%%%%%%%%%%%%%%%%%%%%%

%%%%%%%%%%%%%%%%% BODY OF PAPER %%%%%%%%%%%%%%%%%%

\section{Introduction}

Gamma-ray bursts (GRBs) are the most energetic explosions in the universe. 
It is widely believed that GRBs originate from the death of massive stars \citep{doi:10.1146/annurev.astro.43.072103.150558, Cano2017}  or the merger of two compact objects, such as two neutron stars (NS-NS) \citep{10.1093/mnras/270.3.480, 10.1111/j.1365-2966.2011.18280.x}, one neutron star and one stellar-mass black hole (NS-BH) \citep{Narayan1992}, or one neutron star and one white dwarf  \citep{Zhong_2023}.
There are mainly two phases in GRBs \citep{KUMAR20151}: prompt gamma-ray emission and longer-lasting broadband afterglow emission. 
The prompt emission typically lasts for a timescale of a few milliseconds to a few thousands of seconds 
while the afterglow can persist for a period ranging from days to years. 
According to the internal-external shock model \citep{10.1093/mnras/287.1.110, piran2005physics}, the prompt emission is produced by internal shocks within relativistic outflows, and the afterglow is produced by external shocks between this outflow and its circum-burst medium (CBM).

The external shock model describes the interaction of a relativistic GRB jet with its circum-burst medium. 
This interaction naturally gives rise to a pair of shocks \citep{10.1046/j.1365-8711.1999.02800.x, Sari_1999, Kobayashi_2003, Shao_2005}. 
A long-lived forward shock (FS) sweeps up the CBM, while a short-lived reverse shock (RS) propagates into the jet and sweeps through the jet itself. 
Both forward and reverse shocks can accelerate electrons through stochastic processes \citep{blandford1987particle}, and these non-thermal electrons release their energy through synchrotron radiation and inverse-Compton scattering \citep{Sari_1998, Sari_2001}.
The dynamics and radiation processes of this FS-RS system have been widely studied. 
\citet{Sari_1995} proposed the standard FS-RS model.
\citet{10.1046/j.1365-8711.1998.01681.x}, \citet{Chevalier_1999} and \citet{Wu_2003} investigated the FS-RS system in a stellar wind environment, and obtained  analytical solutions in the cases of an ultra-relativistic reverse shock (RRS) and a Newtonian reverse shock (NRS). \citet{Beloborodov_2006} developed the mechanical model by applying conservation laws for the energy-momentum tensor and the mass flux to the FS-RS system.
\citet{Zhang_Ze_lin_2022} presented a semi-analytical solution for the FS-RS system by considering a power-law density profile of CBM.
In this paper, we utilize the semi-analytical solution proposed by \citet{Zhang_Ze_lin_2022} to describe the evolution of RS. 
Once the RS crosses the shell, we employ a generic dynamical model proposed by \citet{Huang_1999} can describe the evolution of the FS. 

The investigation of off-axis jets in the context of GRBs has been extensive \citep{Woods_1999, Granot_2005, Lamb_2017, Gill_2018, Beniamini_2020}, but the consideration of RS emission from such off-axis jets has been relatively rare.  
\citet{Fraija_2019} conducted a modeling study on the off-axis emission of GW 170817/GRB 170817A and demonstrated that the $\gamma$-ray flux could be consistent with a synchrotron self-Compton (SSC) RS model. 
\citet{Lamb_2019} performed calculations on the RS emission from a structured jet observed from an off-axis perspective, and suggested that the pre-peak afterglow exhibits a distinctive feature originating from the RS. 
In this paper, we calculate the emission of the FS-RS system from a top-hat jet viewed off-axis in a stratified environment, and describe the multi-wavelength observations in GRBs which were previously modeled by an off-axis emission.

This paper is organized as follows. 
In Section~\ref{sec2.1}, we provide an overview of the dynamics and radiation process of relativistic GRB jets. 
In Section~\ref{sec2.2}, we present the calculations of the emission from an off-axis relativistic jet. 
In Section~\ref{sec3}, we apply our model to fit the data from off-axis short GRBs and present the fitting results. 
Finally, Section~\ref{sec4} contains our discussion and conclusions based on findings of this study.

\section{The Model}
    \label{sec2}
\subsection{Dynamics and emission mechanisms of GRB outflows}
\label{sec2.1} % used for referring to this section from elsewhere

We assume a cold shell ejected from a central engine with an isotropic-euqivalent energy $E_\mathrm{iso}$ and an initial Lorentz factor $\Gamma_0$. Thus its mass is $m_\mathrm{ej}=E_\mathrm{iso}/\Gamma_0 c^2$, and it collides with a cold CBM.
The CBM number density is parametrized as $n_1(R)=AR^{-k}$.
Note that $k=0$ and $k=2$ correspond to the ISM and the wind environment respectively, where for the ISM case, $A=n_0$, with the typical value $n_0 = 1\,\mathrm{cm}^{-3}$.
When an FS and an RS are formed, the interaction is described by the FS, contact discontinuity (CD) and RS, which lead to four regions \citep{Sari_1995}: (1) the unshocked CBM, (2) the shocked CBM, (3) the shocked ejecta, and (4) the unshocked ejecta. 
The shocked region is assumed to have a uniform bulk Lorentz factor, that is, $\Gamma_2=\Gamma_3=\Gamma$ (hereafter, the single subscript index $i$ denotes the quantities in Region $i$, and the double index denotes the relative value between two regions).
According to energy conservation and shock jump conditions, we have \citep{Pe'er_2012, Zhang_Ze_lin_2022}
\begin{equation}
    \begin{aligned}
        m_2c^2 + \Gamma_0 m_\mathrm{ej}c^2 =& \Gamma m_2 c^2 + \Gamma(\Gamma-1)m_2 c^2 \\
        &+ \Gamma m_3c^2 + \Gamma(\Gamma_{34}-1)m_3c^2 + \Gamma_0(m_{\mathrm{ej}} - m_3)c^2,
    \end{aligned}
    \label{eq_dyn}
\end{equation}
where $m_2 = 4\pi m_p A R^{3-k}/(3-k) $ is the mass swept by the FS.
The right-hand side of equation (1) is the total energy (including rest energy, kinetic energy, and thermal energy) in the Region 2, Region 3 and Region 4 respectively, while the left-hand side is the total energy before interaction.

Given that $\Gamma^2\gg 1$, the energy conservation equation can be written by
\begin{equation}
    \Gamma = \Gamma_0 \sqrt{\frac{m_3c^2}{2\Gamma_0m_2c^2 + m_3 c^2}},
        \label{eq_gamma1}
\end{equation}
where the mass of shocked ejecta $m_3$ follows Eq.43 in \citet{Zhang_Ze_lin_2022}. 
In the thin-shell case, the evolution of $m_3$ can be reduced into (Eq.47 in \citet{Zhang_Ze_lin_2022})
\begin{equation}
    m_3(R) = \sqrt{\frac{32E_\mathrm{iso}m_pAR^{3-k}}{3(3-k)^2c^2}}.
        \label{eq_m3}
\end{equation}

When $m_3=m_\mathrm{ej}$, the RS crosses the ejecta, and the radius of ejecta shell reaches the crossing radius $R=R_\Delta$ (Eq.48 in Zhang et al.2022). After the RS crosses the shell, energy conservation reads \citep{Huang_1999}
\begin{equation}
    \Gamma^2 m_2 + \Gamma\Gamma_{34,\Delta}m_{\mathrm{ej}} - (\Gamma_\Delta^2 m_{2,\Delta} + \Gamma_\Delta \Gamma_{34,\Delta}m_\mathrm{ej} + m_2 - m_{2,\Delta}) = 0.
        \label{eq_gamma2}
\end{equation}

In order to calculate the synchrotron and inverse-Compton (IC) emission, we assume that electrons in Region 2 and 3 can be accelerated by the FS and RS to a power-law distribution $N_e(\gamma_e)\mathrm{d}\gamma_e\propto\gamma_e'^{-p}\mathrm{d}\gamma_e'$ ($\gamma_e'\ge\gamma_\mathrm{m}' $), and assume $\epsilon_{e,i}$ and $\epsilon_{B,i}$ are the fraction of internal energy in Region $i$ carried by electrons and a magnetic field respectively. 
We use the superscript prime ($'$) to denote the quantities in the rest frame of emitting plasma, the local comoving synchrotron emissivity $P_{\nu'}'$ can be expressed as a broken power law \citep{Gao_2013}. In the slow-cooling regime, 
\begin{equation}
    \frac{P_{\nu'}'}{P_{\nu',\mathrm{max}}'}=\begin{cases}
        (\nu_\mathrm{a}'/\nu_\mathrm{m}')^{1/3}(\nu'/\nu_\mathrm{a}')^2, &\nu'<\nu_\mathrm{a}'<\nu_\mathrm{m}'\\
        (\nu_\mathrm{m}'/\nu_\mathrm{a}')^{(p+4)/2}(\nu'/\nu_\mathrm{m}')^2, &\nu'<\nu_\mathrm{m}'<\nu_\mathrm{a}'\\
        (\nu'/\nu_\mathrm{m}')^{1/3}, &\nu_\mathrm{a}'<\nu'<\nu_\mathrm{m}'\\
        (\nu_\mathrm{a}'/\nu_\mathrm{m}')^{-(p-1)/2}(\nu'/\nu_\mathrm{a}')^{5/2}, &\nu_\mathrm{m}'<\nu'<\nu_\mathrm{a}'\\
        (\nu'/\nu_\mathrm{m}')^{-(p-1)/2}, &\mathrm{max}(\nu_\mathrm{a}',\nu_\mathrm{m}')<\nu'<\nu_\mathrm{c}'\\
        (\nu_\mathrm{c}'/\nu_\mathrm{m}')^{-(p-1)/2}(\nu'/\nu_\mathrm{c}')^{-p/2}, &\nu_\mathrm{c}'<\nu'
    \end{cases}
        \label{eq_syn1}
\end{equation}
and in the fast-cooling regime,
\begin{equation}
    \frac{P_{\nu'}'}{P_{\nu',\mathrm{max}}'}=\begin{cases}
        (\nu_\mathrm{a}'/\nu_\mathrm{c}')^{1/3}(\nu'/\nu_\mathrm{a}')^2, &\nu'<\nu_\mathrm{a}'\\
        (\nu'/\nu_\mathrm{c}')^{1/3}, &\nu_\mathrm{a}'<\nu'<\nu_\mathrm{c}'\\
        (\nu'/\nu_\mathrm{c}')^{-1/2}, &\nu_\mathrm{c}'<\nu'<\nu_\mathrm{m}'\\
        (\nu_\mathrm{m}'/\nu_\mathrm{c}')^{-1/2}(\nu'/\nu_{\mathrm{m}}')^{-p/2}, &\nu_\mathrm{m}'<\nu'
    \end{cases}
        \label{eq_syn2}
\end{equation}
where the first characteristic frequency $\nu_\mathrm{a}'$ is the self-absorption frequency calculated by \citet{Wu_2003},  and $\nu_\mathrm{m}'$, $\nu_\mathrm{c}'$ are characteristic frequencies corresponding to the comoving frame minimum electron Lorentz factor $\gamma_\mathrm{m}'$ and cooling Lorentz factor $\gamma_\mathrm{c}'$. 
The flux normalization and break frequencies are \citep{Gill_2018}
\begin{subequations}
    \begin{align}
        P_{\nu',\mathrm{max},i}'&\simeq \frac{m_e c^2\sigma_T}{3q_e}(8\pi)^{1/2}\epsilon_{B,i}^{1/2}e_i'^{1/2},
            \label{eq_P}\\
        \nu_{\mathrm{m},i}'&=\frac{3\sqrt{2\pi}}{8}\left(\frac{p-2}{p-1}\right)^2\frac{q_e}{m_e^3c^5}\epsilon_{B,i}^{1/2}\epsilon_{e,i}^2 e_i'^{5/2} n_i'^{-2},
            \label{eq_nu_m}\\
        \nu_{\mathrm{c},i}'&=\frac{27\sqrt{2\pi}}{128}\frac{q_em_ec}{\sigma_T^2}\epsilon_{B,i}^{-3/2}e_i'^{-3/2}\left(\frac{\Gamma}{t_\mathrm{lab}}\right)^2.
            \label{eq_nu_c}
    \end{align}
\end{subequations}
In the above equations, $q_e$ is the elementary charge, $\sigma_T$ is the Thomson cross-section, $t_\mathrm{lab}$ is lab frame time, $e'_i$ and $n_i'$ are proper internal energy density and proper electron number density of Region $i$ respectively.

We use the Compton parameter $Y_i = (-1+\sqrt{1+4\xi_e\epsilon_{e,i}\epsilon_{B,i}})/2$ to denote the ratio of the IC luminosity to synchrotron luminosity with $\xi_e$ being the fraction of the electron energy radiated. 
The IC radiation has two parts: synchrotron self-Compton (SSC) radiation and combined-IC radiation, that is, photons from Region $j$ can be scattered by electrons in Region $i$ (where $j=i$ for SSC and $j\ne i$ for combined-IC). 
The IC volume emissivity in the rest frame can be calculated as \citep{Wang_2001} 
\begin{equation}
    j^{\prime \mathrm{IC}}_{\nu',i} = 3\sigma_T \int_{\gamma_{\mathrm{min},i}'}^{\gamma_{\mathrm{max},i}'}\mathrm{d}\gamma_{e,i}'N_{e,i}(\gamma_{e,i}')\int_0^1\mathrm{d}xg(x)\bar{f}_{\nu'_s,j}'(x),
        \label{eq_IC}
\end{equation}
where $x\equiv \nu'/(4\gamma_{e,i}'^2\nu_{s,j}')$, $\bar{f}'_{\nu'_s}$ is the incident-specific synchrotron flux at shock front in the comoving frame, and $g(x)=1+x+2x\ln(x)-2x^2$.

\subsection{Observer-frame lightcurves}
\label{sec2.2}

\begin{figure}
    \centering
    \includegraphics[width=\columnwidth]{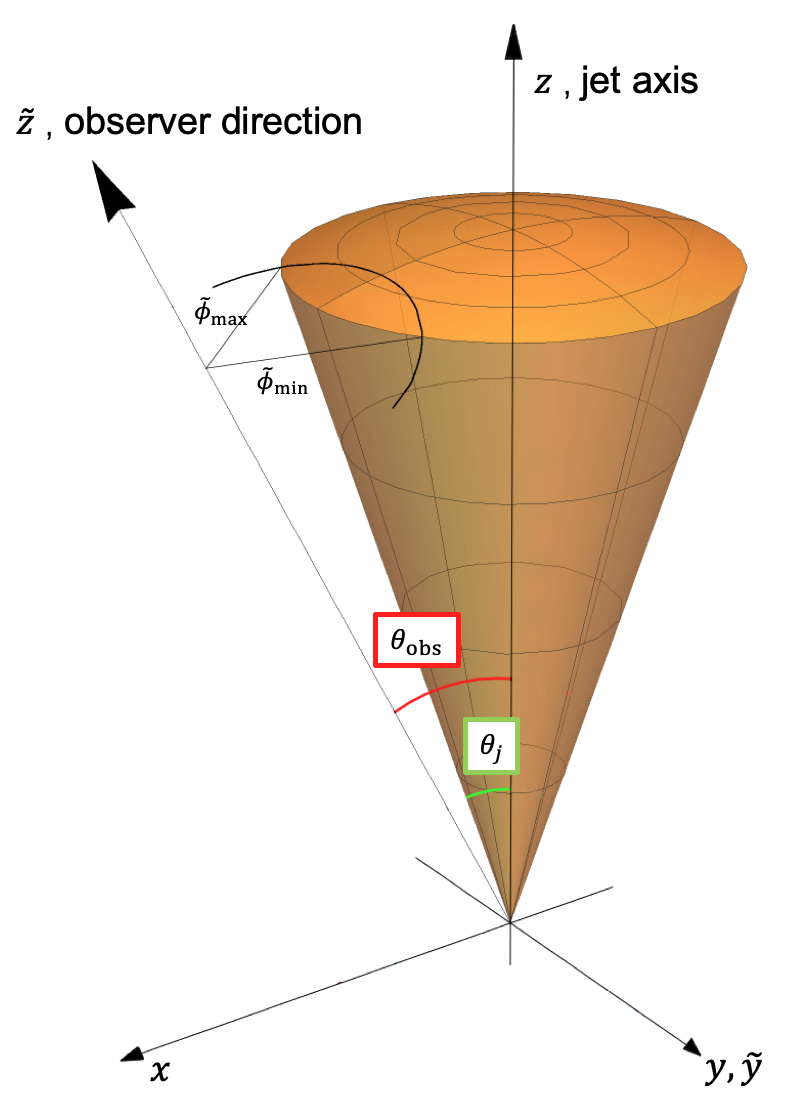}
    \caption{The coordinate system used to calculate afterglow emission from the observer direction. The jet direction is along the $z$ direction, while LOS is along the $\Tilde{z}$ direction. }
    \label{fig1}
\end{figure}

Consider an infinitely thin-shell and ignore the radial structure of emitting regions. The emission originates from polar angles $0\le\theta\le\theta_j$, where $\theta_j$ is the jet half-opening angle. 
We conveniently choose the jet's symmetry axis as $z$ direction, and the direction to the observer, $\hat{n}$, is at an angle $\theta_\mathrm{obs}=\arccos(\hat{n}\cdot \hat{z})$ from the jet axis (see Fig.~\ref{fig1}). 
For a given observed time $t_\mathrm{obs}$, the afterglow emission is obtained by integrating the emissivity over the equal-arrival-time-surface (EATS) which is calculated by \citep{Gill_2018}
\begin{equation}
    t_z\equiv\frac{t_\mathrm{obs}}{1+z}=t_\mathrm{lab}-\frac{R\Tilde{\mu}}{c},
        \label{eq_EATS}
\end{equation}
where $\Tilde{\mu}\equiv \cos\Tilde{\theta}$, $\Tilde{\theta}=\arccos(\hat{r}\cdot\hat{n})$ is an angle from the line-of-sight (LOS), and $z$ is redshift of the source. 
Defining the dimentionless radius $\xi\equiv R/R_\Delta$ (remind $R_\Delta$ is the RS crossing radius), the lab frame time $t_\mathrm{lab}$ depends on the dynamics through the dimentionless velocity $\beta=\sqrt{1-\Gamma^{-2}}$ and can be written as \citep{Gill_2018} 
\begin{equation}
    t_\mathrm{lab}=\frac{R_\Delta}{c}\int_0^\xi \frac{\mathrm{d}\xi'}{\beta(\xi')}.
        \label{eq_t}
\end{equation}

Using the isotropic comoving spectral luminosity $L_{\nu',\mathrm{iso}}'$, the spectral flux can be calculated as \citep{Granot_2005}
\begin{equation}
    F_\nu(t_\mathrm{obs}) = \frac{1+z}{4\pi d_L^2} \int \Tilde{\delta}_D^3 \frac{L_{\nu',\mathrm{iso}}'(\xi)}{4\pi}\mathrm{d}\Tilde{\Omega}, 
        \label{eq_flux}
\end{equation}
where $d_L$ is the luminosity distance, $\Tilde{\delta}_D=[\Gamma(1-\beta\Tilde{\mu})]^{-1}$ is the Doppler factor and $\mathrm{d}\Tilde{\Omega}=\mathrm{d}\Tilde{\mu}\mathrm{d}\Tilde{\phi}$ is the differential solid-angle. 
The integration in the above equation must be calculated over the EATS for a given $t_\mathrm{obs}$. 
Fixing $t_\mathrm{obs}$, one has \begin{equation}
    \mathrm{d}\Tilde{\mu}= \frac{1-\beta\Tilde{\mu}}{\beta\xi}\mathrm{d}\xi.
        \label{eq_J}
\end{equation}
The integration in Eq.~\ref{eq_flux} can be performed over $\xi\in[\xi_\mathrm{min},\xi_\mathrm{max}]$ by using the above equation. 
The two limits of $\xi$ (i.e., $\xi_\mathrm{min}$ and $\xi_\mathrm{max}$) can be calculated by EATS (Eq.~\ref{eq_EATS}), that is, for $\Tilde{\mu}=\Tilde{\mu}_\mathrm{min}=\cos(\theta_\mathrm{obs}+\theta_j)$, $\xi=\xi_\mathrm{min}$, and for $\Tilde{\mu}=\Tilde{\mu}_\mathrm{max}=\cos(\theta_\mathrm{obs}-\theta_j)$, $\xi = \xi_\mathrm{max}$.

In order to integrate over the azimuthal angle $\Tilde{\phi}$, for simplicity and without loss of generality, the LOS is considered to lie in the $xOz$ plane (i.e., $\phi=0$). 
Considering coordinate transformation from $(x,y,z)$ to $(\Tilde{x},\Tilde{y},\Tilde{z})$, we have
\begin{equation}
    \cos\Tilde{\phi} = \frac{\Tilde{\mu}\cdot\cos\theta_\mathrm{obs}-\cos\theta}{\sqrt{1-\Tilde{\mu}^2}\cdot\sin\theta_\mathrm{obs}},
\end{equation}
where $\theta$ is the polar angle in $(x,y,z)$. Fixing $\Tilde{\theta}$ (equivalent to fixing $\Tilde{\mu}$ or $\xi$), the integral over $\Tilde{\phi}$ is (see Fig.~\ref{fig1})
\begin{equation}
    \int_{\Tilde{\phi}_{\min}}^{\Tilde{\phi}_{\max}}\mathrm{d}\Tilde{\phi} = 2\pi - 2\arccos\left( \frac{\Tilde{\mu}\cdot\cos\theta_\mathrm{obs}-\cos\theta_j}{\sqrt{1-\Tilde{\mu}^2}\cdot\sin\theta_\mathrm{obs}}\right).
\end{equation}
For an on-axis observer and $\Tilde{\theta}\le\theta_j-\theta_\mathrm{obs}$, the integral over $\Tilde{\phi}$ is equal to $2\pi$.

\section{Results}
    \label{sec3}
\begin{figure*}
    \centering
    \includegraphics[width=1.0\textwidth, angle=0]{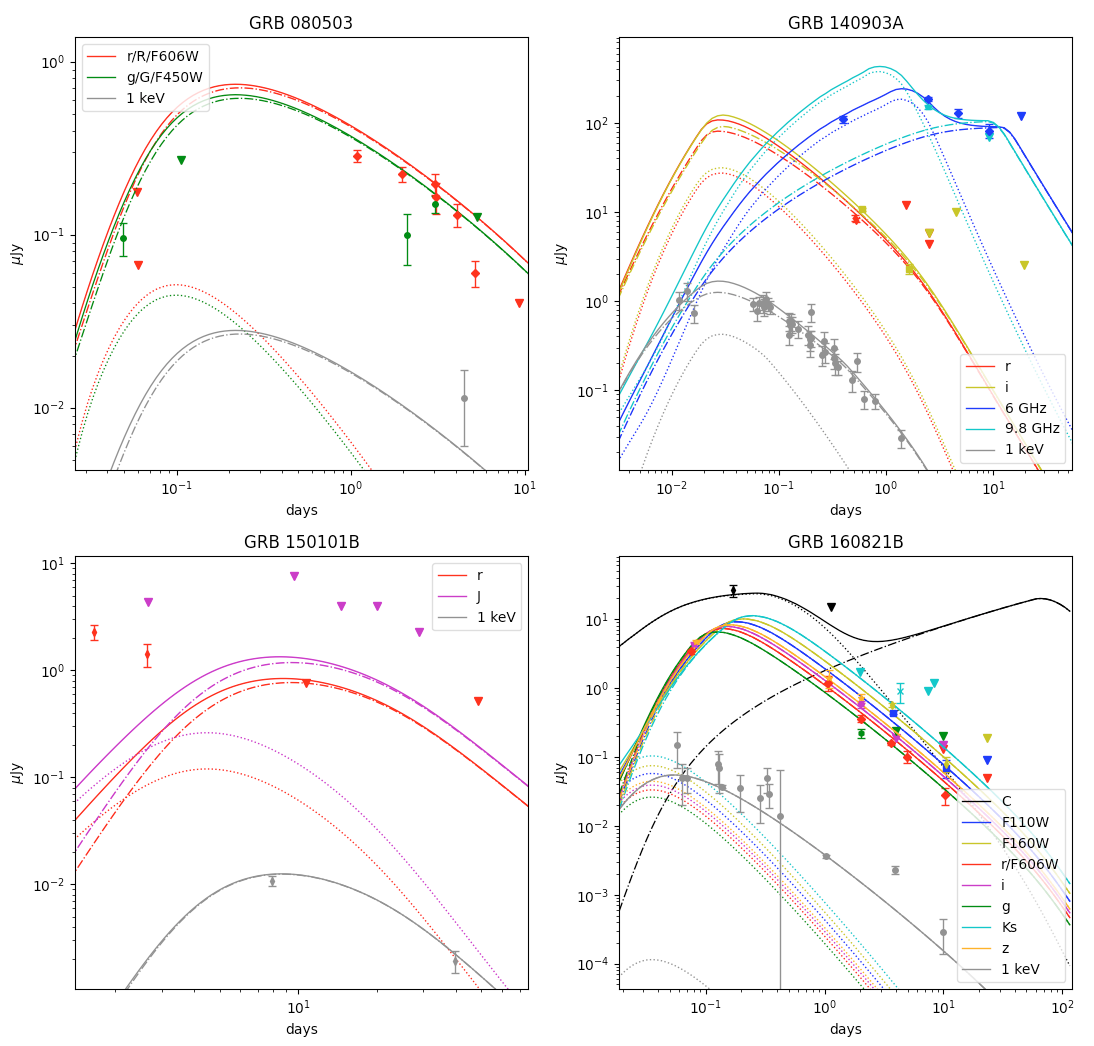}
    \caption{The multi-wavelength afterglow light curves of short GRBs with the best-fit curves in our afterglow model. 
    Different colors are used to represent different bands. 
    The RS lightcurves are depicted as dotted lines, while the FS lightcurves are represented by dash-dotted lines. 
    The multi-wavelength afterglow data for the following GRBs were sourced from the respective literature: GRB 080503 from \citet{Perley_2009}, GRB 140903A from \citet{Fong_2015} and \citet{Troja_2016}, GRB 150101B from \citet{Troja2018} and GRB 160821B from \citet{Troja_2019}. }
    \label{fig_fitting}
\end{figure*}

Due to thin-shell approximation in Eq.~\ref{eq_m3}, we restrict our analysis to the afterglows of short GRBs observed from an off-axis perspective (GRB 080503, GRB 140903A, GRB 150101B, GRB 160821B and GRB 170817A). 
Short GRBs are typically associated with the merger of two compact stars, and their circumburst medium (CBM) is commonly assumed to be a constant density medium. 
In our fitting process, we set $k=0$ to account for this assumption. 
Being consistent with previous studies \citep{ZKP03,Harrison_2013,Gao_2015}, we assume that the electron equipartition $\epsilon_e$ and the electron power-law index $p$ are the same for both the forward-shocked and reverse-shocked regions. 
However, since the outflow may have initial magnetic fields originating from the central engine, we introduce the assumption that the two shock regions have different magnetic equipartition parameters $\epsilon_B$.

\begin{figure}
    \centering
    \includegraphics[width=\columnwidth]{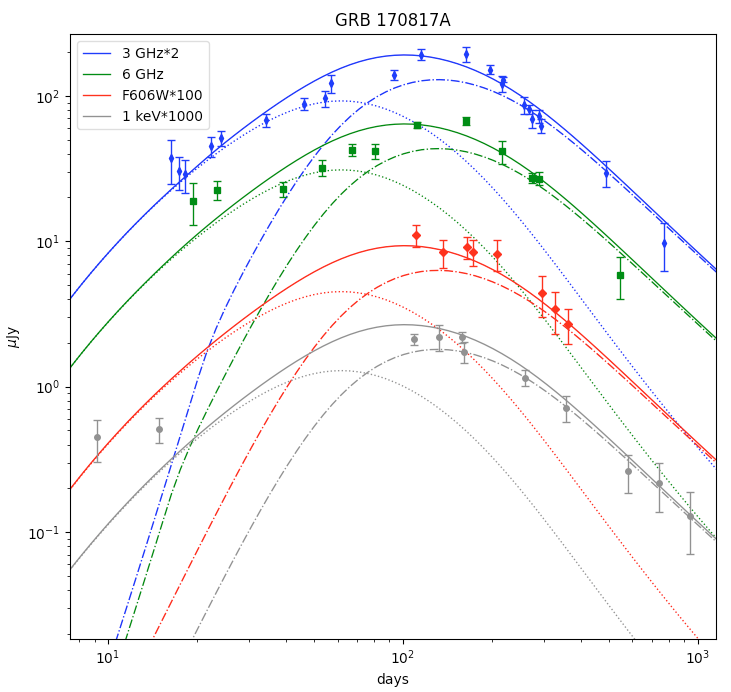}
    \caption{Same as Fig.~\ref{fig_fitting} but only for GRB 170817A. The multiwavelength afterglow data are taken from \citet{Makhathini_2021}. }
    \label{fig_170817A}
\end{figure}

We use Markov Chain Monte Carlo (MCMC) method to determine the best-fit values of the parameters that characterize the multi-wavelength afterglow emission observations with some upper limits. A set of nine parameters, $\{ p, n_1, \theta_\mathrm{obs}, \theta_j, \Gamma_0, E_\mathrm{iso}, \epsilon_e, \epsilon_{B,2}, \epsilon_{B,3} \}$, are required. 
In this study, we adopt uniform priors for the following parameters: $p$, $\log n_1$, $\theta_\mathrm{obs}$, $\theta_j$, $\log \Gamma_0$, $\log E_\mathrm{iso}$, $\log \epsilon_e$, $\log\epsilon_{B,2}$ and $\log \epsilon_{B,3}$.
The specific range for these priors are provided in Table \ref{tab:prior}.
For GRB 170817A, \citet{Abbott_2019} derived the binary inclination angle $\theta_{JN}={151_{-11}^{+15}}^\circ$ by jointing gravitational-wave and electromagnetic observations.
Therefore, we constrain the prior range for $\theta_\mathrm{obs}$ to $0.23\le \theta_\mathrm{obs} \le 0.7$.
Considering that some GRBs may be slight off-axis with limited observational data, making characteristics of off-axis emission less apparent, we restrict our model to an off-axis scenarios, meaning that $\theta_\mathrm{obs}>\theta_j$.
Figure~\ref{fig_fitting} and Figure~\ref{fig_170817A} displays the observations and upper limits for X-ray, optical, and radio data, along with the fitting results obtained using our model. The best-fit values are presented in Table \ref{tab:fitting_parameters}.

\begin{table*}
        \renewcommand\arraystretch{1.5}
	\centering
	\caption{The range for the priors.}
	\label{tab:prior}
	\begin{tabular}{lccccc} % 5 columns, alignment for each
		\hline
		Parameters  & GRB 080503 & GRB 140903A & GRB 150101B & GRB 160821B & GRB 170817A\\
		\hline
		$p$                           &(2, 3) &(2, 3) &(2, 3) &(2,3) &(2, 3)\\
            $\log[n_1 (\mathrm{cm}^{-3})]$   &(-5, 0.5)  &(-5, 0.5) &(-5, 0.5) &(-5, 0.5) &(-5, 0.5)\\
            $\theta_\mathrm{obs}$ (rad)     &(0, 0.7) &(0, 0.7) &(0, 0.7) &(0, 0.7) &(0.23, 0.7)\\
            $\theta_j$\quad (rad)           &(0, 0.2) &(0, 0.5)  &(0, 0.5) &(0, 0.2) &(0, 0.5)\\
            $\log \Gamma_0$                 &(1, 3) &(1, 3) &(1, 3) &(1, 3) &(1, 3)\\
		$\log[E_\mathrm{iso} (\mathrm{erg})]$  &(49, 54) &(49, 54) &(49, 54) &(49, 54) &(49, 54)\\
            $\log \epsilon_e$               &(-6, -0.5) &(-6, -0.5) &(-6, -0.5) &(-6, -0.5) &(-6, -0.5)\\
            $\log \epsilon_{B,2}$             &(-6, -0.5) &(-6, -0.5) &(-6, -0.5) &(-6, -0.5) &(-6, -0.5)\\
            $\log \epsilon_{B,3}$           &(-6, -0.5) &(-6, -0.5) &(-6, -0.5) &(-6, -0.5) &(-6, -0.5)\\
		\hline
	\end{tabular}
\end{table*}

\begin{table*}
        \renewcommand\arraystretch{1.5}
	\centering
	\caption{Fitting parameters used to describe the multi-wavelength afterglow lightcurves.}
	\label{tab:fitting_parameters}
	\begin{tabular}{lccccc} % 5 columns, alignment for each
		\hline
		Parameters  & GRB 080503 & GRB 140903A & GRB 150101B & GRB 160821B & GRB 170817A\\
		\hline
		$p$                                    & $2.089_{-0.062}^{+0.117}$  & $2.318_{-0.069}^{+0.047}$  &$2.317_{-0.022}^{+0.009}$  &$2.824_{-0.025}^{+0.028}$  &$2.152_{-0.005}^{+0.006}$\\
            $\log[n_1 (\mathrm{cm}^{-3})]$         & $-1.027_{-1.716}^{+1.377}$ & $-1.490_{-0.407}^{+0.516}$ &$-1.565_{-0.819}^{+0.639}$ &$-4.740_{-0.174}^{+0.288}$ &$-3.091_{-0.128}^{+0.188}$ \\
            $\theta_\mathrm{obs}$ (rad)            & $0.183_{-0.052}^{+0.035}$  & $0.074_{-0.014}^{+0.028}$  &$0.155_{-0.026}^{+0.021}$  &$0.152_{-0.033}^{+0.038}$  &$0.250_{-0.007}^{+0.021}$ \\
            $\theta_j$\quad (rad)                  & $0.150_{-0.049}^{+0.032}$  & $0.065_{-0.014}^{+0.030}$  &$0.054_{-0.010}^{+0.009}$  &$0.139_{-0.033}^{+0.038}$  &$0.067_{-0.007}^{+0.008}$ \\
            $\log \Gamma_0$                        & $2.459_{-0.448}^{+0.340}$  & $1.983_{-0.213}^{+0.293}$  &$2.419_{-0.148}^{+0.108}$  &$2.748_{-0.230}^{+0.163}$  &$2.638_{-0.073}^{+0.065}$ \\
		$\log[E_\mathrm{iso} (\mathrm{erg})]$  & $52.426_{-0.907}^{+0.863}$ & $51.530_{-0.228}^{+0.242}$ &$53.329_{-0.600}^{+0.424}$ &$50.637_{-0.127}^{+0.180}$ &$53.333_{-0.125}^{+0.101}$\\
            $\log \epsilon_e$                      & $-1.529_{-1.054}^{+0.697}$ & $-0.626_{-0.124}^{+0.091}$ &$-1.547_{-0.348}^{+0.484}$ &$-0.540_{-0.044}^{+0.029}$ &$-0.584_{-0.116}^{+0.061}$ \\
            $\log \epsilon_{B,2}$                  & $-3.829_{-1.320}^{+1.422}$ & $-2.467_{-0.700}^{+0.351}$ &$-3.601_{-0.525}^{+0.746}$ &$-0.865_{-0.281}^{+0.244}$ &$-6.899_{-0.072}^{+0.139}$ \\
            $\log \epsilon_{B,3}$                  & $-2.741_{-1.464}^{+1.523}$ & $-1.724_{-0.835}^{+0.840}$ &$-1.423_{-0.802}^{+0.603}$ &$-1.090_{-0.553}^{+0.383}$ &$-3.377_{-0.137}^{+0.117}$ \\
		\hline
	\end{tabular}
\end{table*}

\textit{GRB 080503} is a short GRB that exhibits a late rebrightening feature. 
This feature has been proposed by two models: a magnetar-powered merger-nova \citep{Gao_2015_080503} and an off-axis jet scenario where a jet becomes visible once relativistic beaming is reduced by deceleration \citep{Perley_2009, Fraija_2022}. 
The multi-wavelength afterglow observations, together with the fits computing using the FS-RS model for an top-hat jet model, are shown in top left panel of Fig.~\ref{fig_fitting}. 
Additionally, Fig.~\ref{corner080503} displays the corner plots showing the results of our MCMC parameter estimation.
The fitting results indicate that both the optical and X-ray afterglows are primarily dominated by FS emission. 
The best-fit values obtained from the fitting process suggest that the jet is slightly off-axis, with a viewing angle $\theta_\mathrm{obs} = {10.49_{-2.98}^{+2.01}}^\circ$ and a half-opening angle $\theta_j = {8.59_{-2.81}^{+1.83}}^\circ$. 
Furthermore, the best-fit values for the isotropic energy $E_\mathrm{iso}\approx 2.67\times10^{52}$erg and the initial bulk Lorentz factor $\Gamma_0\approx287.74$ indicate that the jet is narrowly collimated. 
Additionally, measurements of the linear polarization of the optical afterglow could potentially valuable insights to distinguish our off-axis scenario from the on-axis scenario \citep{Pedreira_2023}.

\textit{GRB 140903A} is a nearby (redshift $z=0.351$) short GRB. 
The afterglow was reported and analyzed by \citet{Troja_2016} for the first 2 weeks.
They found that this burst was caused by a collimated jet viewed off-axis and also originated from the merger of two compact objects.
In this paper, we use the FS-RS model for an off-axis top-hat jet to fit the observations.
In the top right panel of Fig.~\ref{fig_fitting}, we present the multi-wavelength observations of its afterglow, along with the fits derived by using the FS-RS model for an off-axis jet. 
Fig.~\ref{corner140903A} exhibits the best-fit values and the median of the posterior distributions of the parameters.
In the optical and X-ray bands, the RS emission is weaker than the FS emission. 
However, it is possible that the RS contribution becomes significant in the radio waveband.
The best-fit values obtained from the fitting process reveal a slightly off-axis relativistic jet, with a viewing angle of  $\theta_\mathrm{obs}={4.24_{-0.80}^{+1.60}}^\circ$ and a half-opening angle of $\theta_j = {3.72_{-0.80}^{+1.72}}^\circ$.
The fitting results also indicate that the jet was narrowly collimated, with an estimated isotropic energy of $E_\mathrm{iso} \approx 3.39\times 10^{51}$ erg and an initial bulk Lorentz factor of $\Gamma_0 \approx 96.16$.

\textit{GRB 150101B}, 
a short GRB similar to GRB 170817A, is located at redshift of $z=0.1341$.
The optical emission observed around $\sim 2$ days is powered by a blue kilonova, and the long-live X-ray afterglow suggests that the GRB afterglow was viewed off-axis \citep{Troja2018}.
The multi-wavelength observations and the fits derived in our model are displayed in the bottom left panel of Fig.~\ref{fig_fitting}. 
Fig.~\ref{corner150101B} presents the corner plots that depict the best-fit values and the median of the posterior distribution of the parameters.
At early times, the optical afterglow is significantly fainter than the kilonova emission, and potential RS emission is not observed.
The absence of X-ray emission during the early times can be attributed to an off-axis jet.
The best-fit values of the viewing angle $\theta_\mathrm{obs}={8.88_{-1.49}^{+1.20}}^\circ$ and half-opening angle $\theta_j={3.09_{-0.57}^{+0.52}}^\circ$ indicate a mildly inclined jet similar to GRB 170817A. The relatively higher redshift makes it more challenging to observe.
Furthermore, the best-fit values for the isotropic energy ($E_\mathrm{iso}\approx 2.13\times10^{53}$ ergs) and initial Lorentz factor ($\Gamma_0\approx262$) imply that the afterglow emission originates from a narrowly collimated jet.

\textit{GRB 160821B} is a short GRB at redshift $z=0.162$. 
In the bottom right panel of Fig.~\ref{fig_fitting}, we present the multi-wavelength observations starting from $\sim 0.1$ day after the bursts, along with the fits derived in our model. 
Fig.~\ref{corner160821B} displays the corner plots showing the results of our MCMC parameter estimation.
In the study of GRB 160821B conducted by \citet{Lamb_2019_160821B}, the reverse shock is dominant at $\lesssim 1$ day at 5 GHz, a kilonova component emerged in the optical and infrared bands during a time interval of $\sim$ 1-5 days. 
Additionally, an excess in X-ray at 3-4 days suggests the presence of energy injection.
In our model, the RS emission dominates C-band observation at early times, which is consistent with the analysis by \citet{Lamb_2019_160821B}.
The best-fit values obtained from our analysis reveal a slightly off-axis jet, with a viewing angle of $\theta_\mathrm{obs} = {8.71_{-1.89}^{+2.18}}^\circ$ and a half-opening angle of $\theta_j={7.96_{-1.89}^{+2.18}}^\circ$ suggest that a slightly off-axis jet, which are comparable with \citet{Jin_2018_160821B}. 
Furthermore, the best-fit values for the isotropic energy $E_\mathrm{iso}\approx 4.34\times10^{50}$ erg and the initial Lorentz factor $\Gamma_0 \approx 559.76$ indicate that the jet was narrowly collimated.

\textit{GW 170817/GRB 170817A} is notable as the first binary neutron star merger detected in gravitational waves, and uniquely, it was also accompanied by electromagnetic radiation. 
Several explanations for the afterglow has been proposed, including a structured jet viewed off-axis \citep{Gill_2018, Lamb_2019_170817}, energy injection into a top-hat jet \citep{Li_2018,Yu_2018}, refreshed shock in an off-axis top-hat jet \citep{Lamb2020} and a relativistic electron-positron wind observed off-axis \citep{Li_2021}. 
In Fig.~\ref{fig_170817A}, we present the multi-wavelength observations of the afterglow, along with the fitting results obtained using the FS-RS model for an off-axis top-hat jet. 
Fig.~\ref{corner170817A} exhibits the best-fit values and the median of the posterior distributions of the parameters.
In particular, our model can reproduce the lightcurves without invoking a structured jet or energy injection.
The fitting results suggest that RS emission may dominate the observations and reaches its peak before $\sim10^2$ days.
The best-fit value of the viewing angle is $\theta_\mathrm{obs} = {14.32_{-0.40}^{+1.20}}^\circ$, 
closely approaching the upper limit of the binary inclination angle determined by \citet{Abbott_2019}.
Furthermore, the best-fit values of the jet's half-opening angle $\theta_j = {3.84_{-0.40}^{+0.46}}^\circ$, isotropic energy $E_\mathrm{iso}\approx 2.15\times10^{53}$erg and initial bulk Lorentz factor $\Gamma_0\approx 434.51$ suggest that synchrotron afterglow emission originates from a narrowly collimated jet.
The best-fit value of the constant-density CBM is $n_1 = 8.11_{-2.07}^{+4.39}\times 10^{-4}~\mathrm{cm}^{-3}$, indicating that GW 170817/GRB 170817A occurred in an environment with a very low density, similar to many other short GRBs.

The superluminal motion of the flux centroid was observed in GRB 170817A \citep{Mooley2018,Mooley2022}.
To calculate the proper motion of flux centroid, we consider the projected image of the outflow on plane $\Tilde{x}O\Tilde{y}$ with coordinates $(\Tilde{x},\Tilde{y})$.
In this plane, the line connecting the LOS to the jet axis coincides with the $\Tilde{x}$-axis (see Fig.~\ref{fig1}).
Due to symmetry, the flux centroid will move exclusively along the $\Tilde{x}$-axis. Its location can be expressed as
\begin{equation}
    \Tilde{x}_c(t_\mathrm{obs})=\frac{\frac{1+z}{4\pi d_L^2} \int \Tilde{\delta}_D^3 L_{\nu'}' R\cos\Tilde{\theta}\cos\Tilde{\phi}~\mathrm{d}\Tilde{\Omega}}{F_{\nu}(t_\mathrm{obs})}.
\end{equation}
The angular displacement $\Tilde{\theta}_c$ from the location of GRB central source is given by $\Tilde{\theta}_c=\Tilde{x}_c/d_A$, with $d_A$ representing the angular diameter distance. We use $d_A=40\,$Mpc for GRB 170817A. 
Finally, we can calculate the average apparent velocity of flux centroid over a time interval $[t_{\mathrm{obs},i}, t_{\mathrm{obs},j}]$ using the equatio,
\begin{equation}
    v_{\mathrm{app}}=\frac{\Tilde{x}_{c,j}-\Tilde{x}_{c,i}}{t_{\mathrm{obs},j}-t_{\mathrm{obs},i}},
\end{equation}
The dimensionless velocity is given by $\beta_\mathrm{app} = v_\mathrm{app}/c$.
\begin{figure}
    \centering
    \includegraphics[width=\columnwidth]{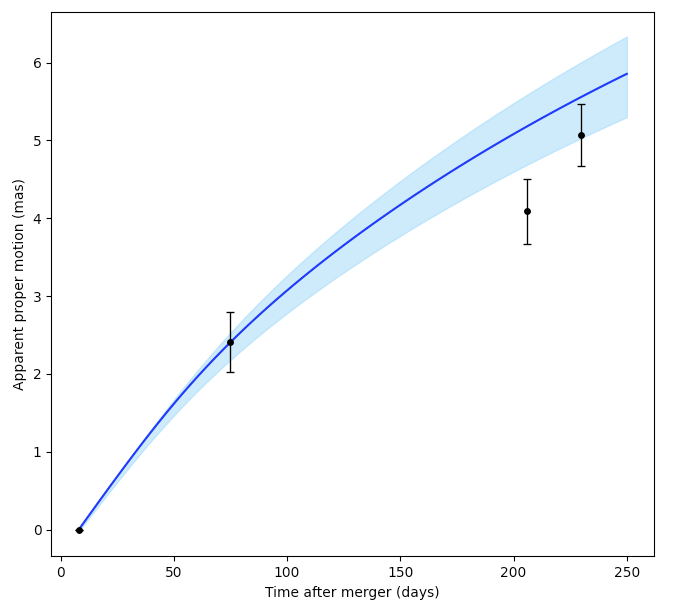}
    \caption{Proper motion of GRB 170817A relative to 8 d Hubble Space Telescope (HST) measurement. The blue line represents the result obtained from our model, based on the best-fitting parameters, while the shaded areas depict the 68\% confidence intervals. The proper motion measured between 8 d - 75 d, 8 d - 206 d, and 8 d - 230 d is $2.41\pm0.38$ mas, $4.09\pm0.42$ mas, and $5.07\pm0.40$ mas, respectively \citep{Mooley2022}. }
    \label{fig_pm}
\end{figure}

In Fig.~\ref{fig_pm}, we present the angular displacement of flux centroid calculated by our model with the fitting parameters. 
The consistency between the model and the data suggests that our model can also reproduce the superluminal motion of the flux centroid.
The average apparent velocities from 8 d to 75 d and from 8 d to 230 d, calculated using the best-fitting parameters, are $\beta_{\mathrm{app,8d-75}}=8.29$ and $\beta_\mathrm{app,8d-230d}=5.78$, respectively, which are comparable with the results ($\beta_{\mathrm{app,8d-75}}=7.6\pm1.3$ and $\beta_\mathrm{app,8d-230d}=5.2\pm0.5$) in \citet{Mooley2022}.

\section{Discussion and conclusions}
    \label{sec4}
We have focused on investigating the contribution of the RS to the afterglow from a top-hat jet viewed off-axis in the thin-shell case. 
According to the standard forward-reverse shock model, the RS is initially non-relativistic for the thin-shell case.
As the shell spreads, the RS may transition to a mildly relativistic \citep{Sari_1995}. 
If a GRB jet is highly magnetized, the RS can be significantly suppressed, and only the FS emission would be observed \citep{ZK05}.
For simplicity, we assume a low degree of magnetization for the GRB jet (i.e., $\sigma \ll 1$), allowing the RS to form soon after the jet interacts with CBM. 
In a more realistic and complex scenario, the jet is likely to be magnetized, and the degree of magnetization is typically $\sigma\lesssim$ a few when the shell is coasting. 
The RS forming radius connects with the degree of magnetization $\sigma$ and a dimensionless quantity $\xi$ \citep{MZ2022}, and the dimensionless quantity $\xi$ is defined by \citet{Sari_1995} 
\begin{equation}
    \xi\equiv (3-k)^{-\frac{1}{2(3-k)}} \left(\frac{l}{\Delta_0}\right)^{1/2}\Gamma_0^{-(4-k)/(3-k)},
\end{equation}
where $l$ is the Sedov length, and $\Delta_0$ is the initial width of shell. 
Consequently, different angular directions within a structured jet will have different RS forming radii. 
We plan to delve into this aspect in detail in a future paper. 
\begin{figure}
    \centering
    \includegraphics[width=\columnwidth]{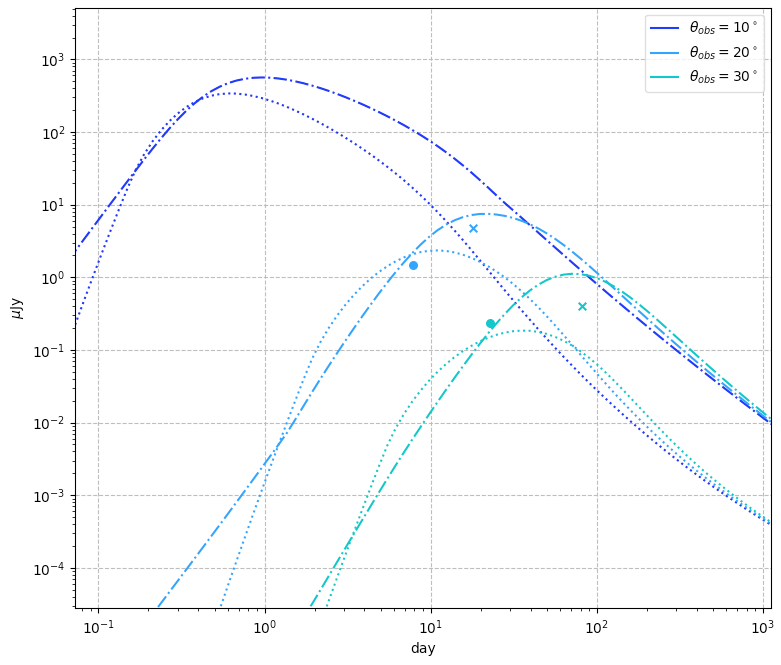}
    \caption{Afterglow R-band lightcurves for jets viewed from different angles. The dotted lines represent the RS lightcurves, while the dash-dot lines represent the FS lightcurves. 
    The dots on the plot are the approximated locations of RS peaks, while the crosses mark the approximated locations of FS peaks. 
    The model parameters used in the calculation are as follows: the half-opening angle of jets is $\theta_j = 6^\circ$, the isotropic energy of jets is $E_\mathrm{iso}=2\times 10^{52}~\mathrm{erg}$, the initial bulk Lorentz factor is $\Gamma_0 = 100$, and CBM density is $n_1 = 0.1~\mathrm{cm}^{-3}$ ($k=0$). }
    \label{fig_LC}
\end{figure}

If the jet can be approximated as a point source (which means $\theta_\mathrm{obs}\gg\theta_j$), we can describe the jet approximately using a single Doppler factor. 
The afterglow lightcurve can then be calculated by
\begin{equation}
    F_\nu(\theta_\mathrm{obs},t) = a^3 F_{\nu/a}(\theta_\mathrm{obs}=0,at) ,
\end{equation}
where $a$ represents the ratio of the Doppler factor for the off-beam observer to that of the on-beam observer. 
The RS emission reaches its peak when the RS crosses the shell, and the peak time of RS emission can be approximated as:
\begin{equation}
    t_{\mathrm{p},r}\approx \left[1 + \Gamma_0^2(\theta_\mathrm{obs}-\theta_j)^2\right] \frac{l}{c\Gamma_0^{2(4-k)/(3-k)}} .
\end{equation}
The FS emission peaks at a time determined by
\begin{equation}
    \theta_\mathrm{obs} - \theta_j = \frac{1}{\Gamma(\xi=\xi_f)} , 
\end{equation}
and substituting $\xi_f$ into the EATS (Eq.~\ref{eq_EATS}) yields the FS peak time $t_{\mathrm{p},f}$. 
The peak flux of FS (for $k=0$) can be calculated as \citep{Nakar_2002,Lamb_2017}
\begin{equation}
    F_{\nu,\mathrm{p},f} \propto E_\mathrm{iso}~n_1^{(p+1)/4} \epsilon_{e}^{p-1}\epsilon_{B,2}^{(p+1)/4}\nu_\mathrm{obs}^{(1-p)/2}d_L^{-2}\theta_\mathrm{obs}^{-2p} \theta_j^2 .
\end{equation}
The approximate values for the FS and RS peaks are shown in Fig.~\ref{fig_LC}. 
Notably, the RS emission has a lower peak compared to the FS emission. 
This is because when the RS emission reaches its peak, the line of sight (LOS) has not yet entered the beaming cone of the jet, resulting in a reduced observed flux. 

In this paper, we have applied the forward-reverse shock model to an off-axis top-hat jet in a stratified environment. 
Based on the model, we fit the multi-wavelength afterglow lightcurves of GRB 080503, GRB 140903A, GRB 150101B, GRB 160821B and GW 170817/GRB 170817A. 
The MCMC method is adopted to obtain the best-fitting parameters. 
According to the fitting parameters, we find that (i) the jets of GRB 080503, GRB 140903A and GRB 160821B are slightly off-axis, while GRB 150101B and gravitational wave(GW)-detected merger GW 170817/GRB 170817A are mildly inclined;
(ii) the density of CBM is relatively low in short GRBs. 
Furthermore, we noted that GRB 170817A, in contrast to the other three GRBs, has a much smaller luminosity distance. 
This smaller distance allows us to detect GRB 170817A from mildly inclined viewing angle. 

Orphan afterglows may arise from mildly inclined GW-detected mergers, and the optical orphan afterglow would be found at early stage, when the afterglow emission still rising. 
At this phase, the reverse shock may play a significant role in shaping the optical emission and produce a distinct feature before the RS peak time $t_{\mathrm{p},r}$, which typically occurs within tens of seconds to $\sim 10$ days after merger. 
The reverse shock component holds potential for constraining the physical parameters of the orphan afterglow and improving our understanding of these events. 

\section*{acknowledgments}
We thank Professor Rui-Zhi Yang for generously providing the computational resources on the server, which has been instrumental in advancing our research. This work was supported by the National SKA Program of China (grant No. 2020SKA0120300) and National Natural Science Foundation of China (grant No. 12393812).

\section*{data availability}
The data developed for the calculation in this work is available upon request.

%\section*{Acknowledgements}

%The Acknowledgements section is not numbered. Here you can thank helpful
%colleagues, acknowledge funding agencies, telescopes and facilities used etc.

%%%%%%%%%%%%%%%%%%%%%%%%%%%%%%%%%%%%%%%%%%%%%%%%%%
%\section*{Data Availability}

%The inclusion of a Data Availability Statement is a requirement for articles published in MNRAS. Data Availability Statements provide a standardised format for readers to understand the availability of data underlying the research results described in the article. The statement may refer to original data generated in the course of the study or to third-party data analysed in the article. The statement should describe and provide means of access, where possible, by linking to the data or providing the required accession numbers for the relevant databases or DOIs.



\bibliographystyle{mnras}
\bibliography{references} % if your bibtex file is called example.bib

%\begin{thebibliography}{99}
%\end{thebibliography}

%%%%%%%%%%%%%%%%%%%%%%%%%%%%%%%%%%%%%%%%%%%%%%%%%%

%%%%%%%%%%%%%%%%% APPENDICES %%%%%%%%%%%%%%%%%%%%%

\appendix
\section{corner plots of the fitting results}

\begin{figure*}
    \centering
    \includegraphics[width=1.0\textwidth, angle=0]{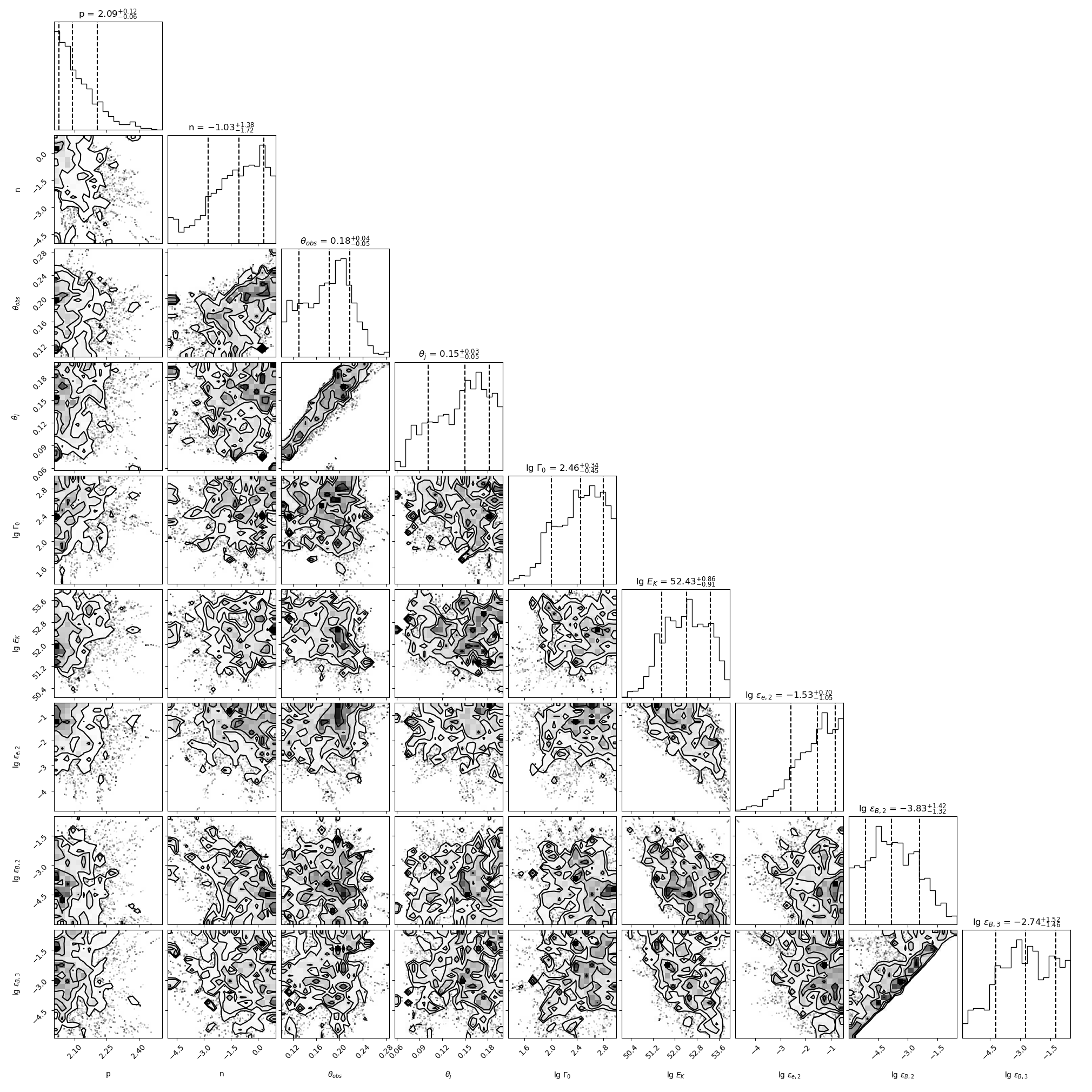}
    \caption{Corner plot of the parameters derived from fitting the multi-wavelength lightcurves of GRB 080503 with our model. Our best-fitting parameters and corresponding 1$\sigma$ uncertainties are shown with the black dashed lines in the histograms on the diagonal.}
    \label{corner080503}
\end{figure*}

\begin{figure*}
    \centering
    \includegraphics[width=1.0\textwidth, angle=0]{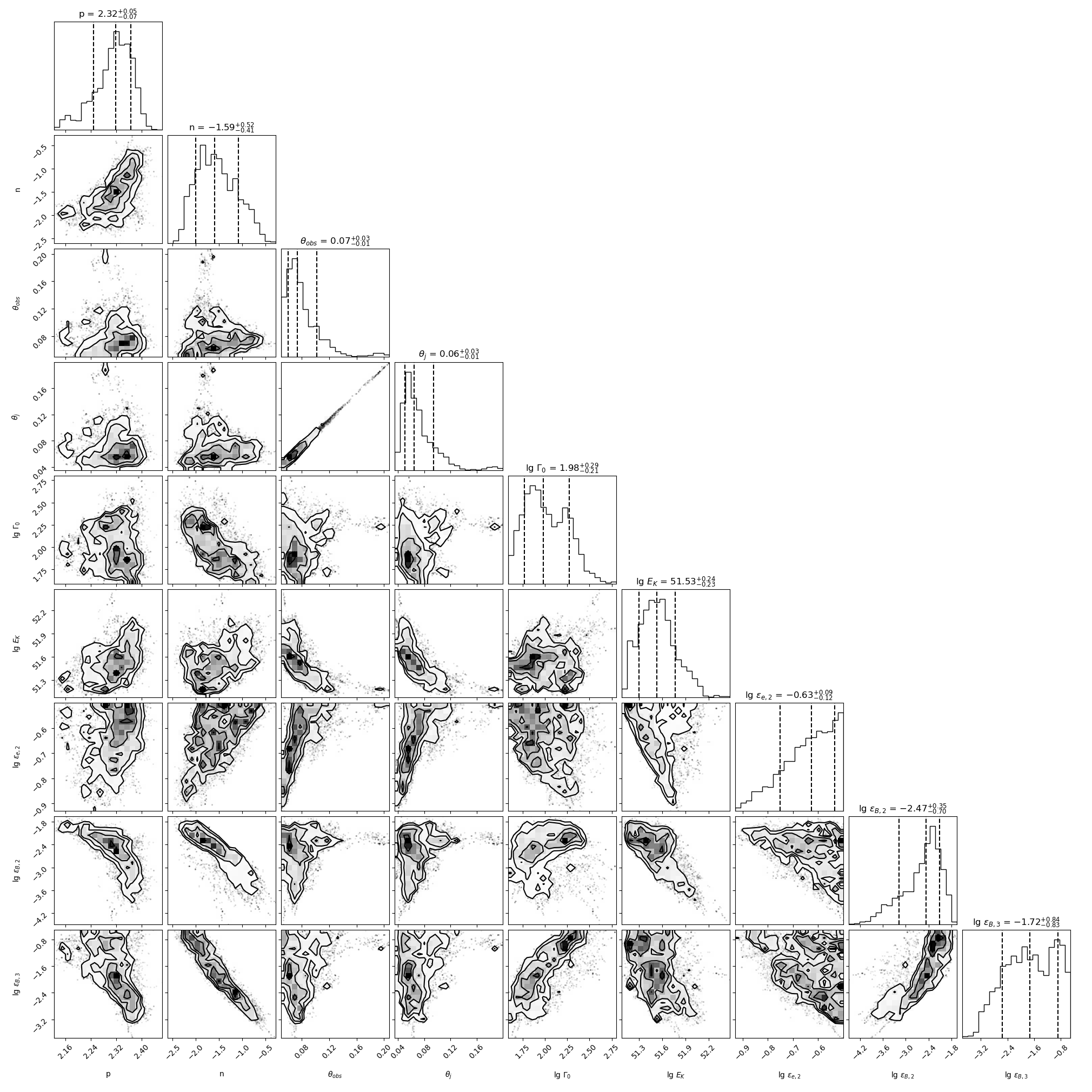}
    \caption{Same as Fig.~\ref{corner080503} but for GRB 140903A.}
    \label{corner140903A}
\end{figure*}

\begin{figure*}
    \centering
    \includegraphics[width=1.0\textwidth, angle=0]{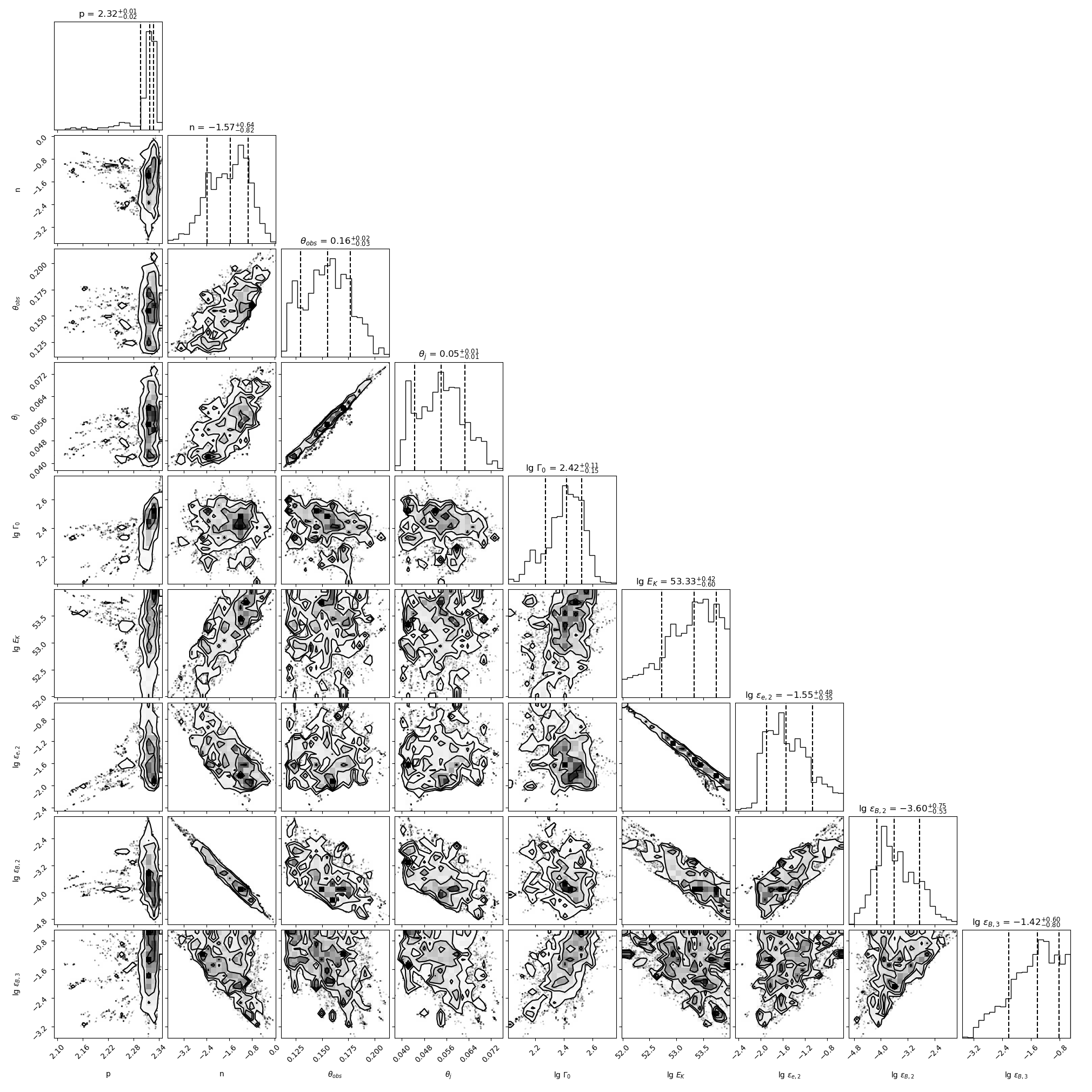}
    \caption{Same as Fig.~\ref{corner080503} but for GRB 140903A.}
    \label{corner150101B}
\end{figure*}

\begin{figure*}
    \centering
    \includegraphics[width=1.0\textwidth, angle=0]{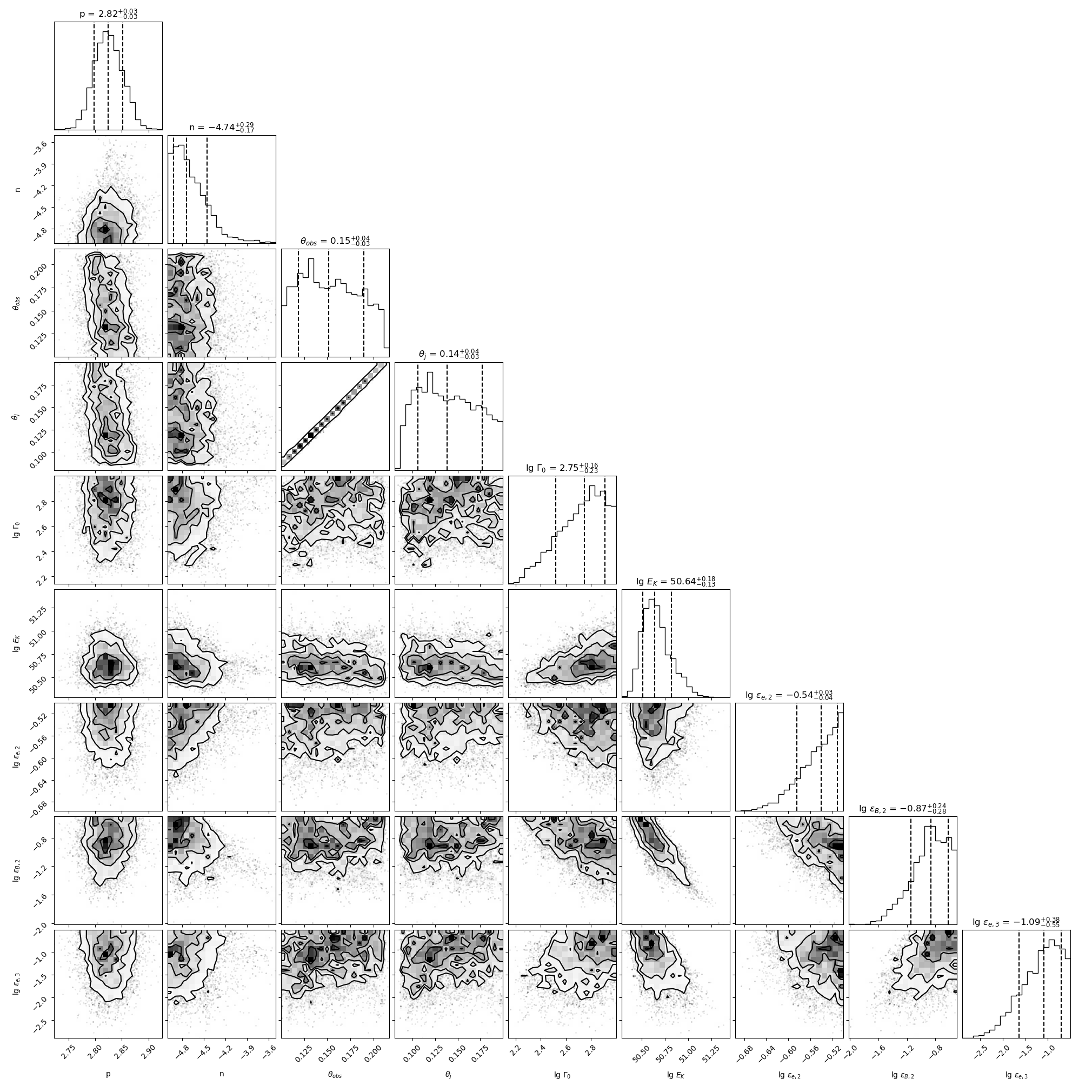}
    \caption{Same as Fig.~\ref{corner080503} but for GRB 160821B.}
    \label{corner160821B}
\end{figure*}

\begin{figure*}
    \centering
    \includegraphics[width=1.0\textwidth, angle=0]{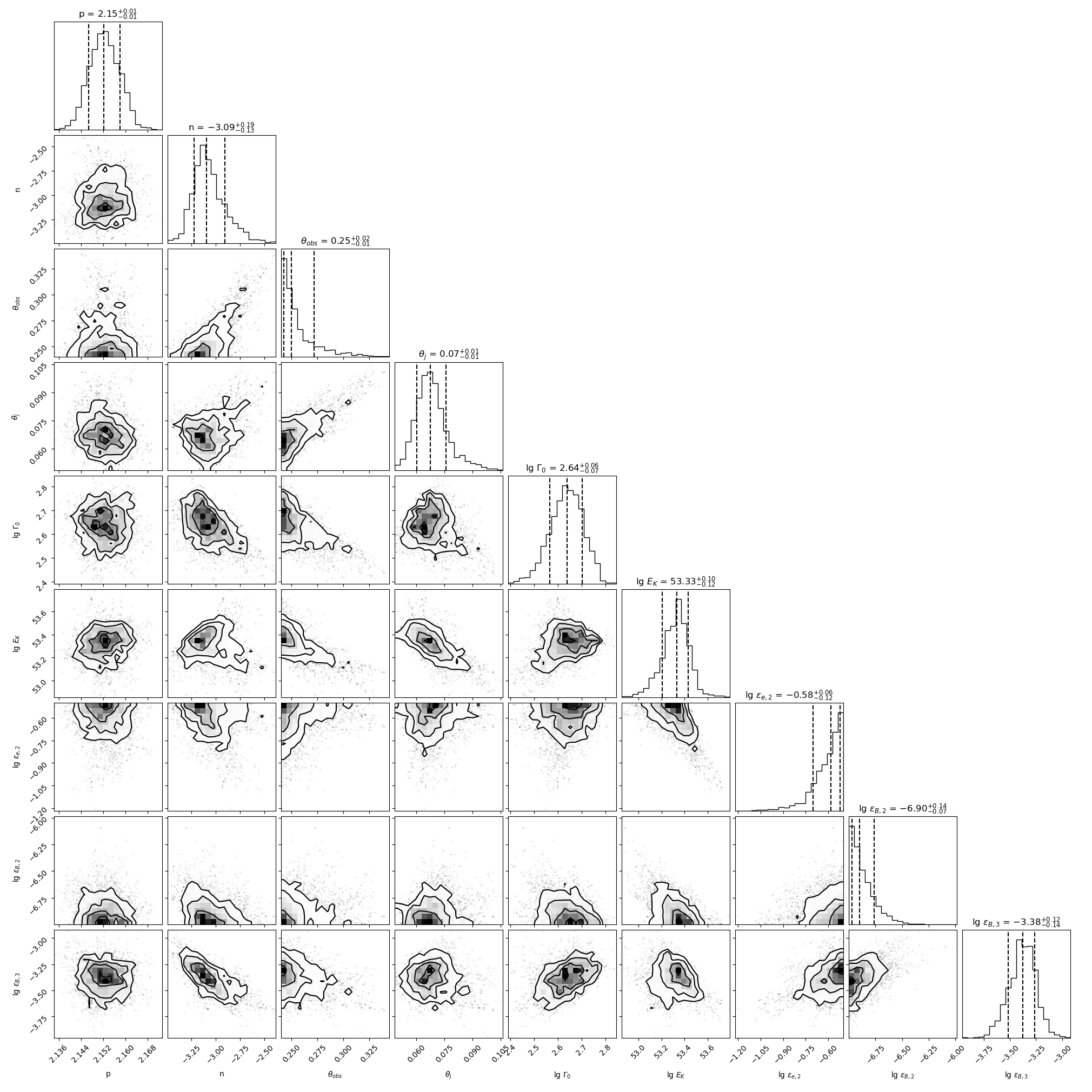}
    \caption{Same as Fig.~\ref{corner080503} but for GRB 170817A.}
    \label{corner170817A}
\end{figure*}

We also show the corner plots of the fitting results in Fig.~\ref{corner080503}-\ref{corner170817A}.

%%%%%%%%%%%%%%%%%%%%%%%%%%%%%%%%%%%%%%%%%%%%%%%%%%

% Don't change these lines
\bsp	% typesetting comment
\label{lastpage}
\end{document}